\documentclass[letterpaper,twocolumn,10pt]{article}
\usepackage{usenix,epsfig,endnotes}
\usepackage{cite}
\usepackage{url}
\usepackage{parskip}
\usepackage{indentfirst} 
\usepackage{color}
\usepackage{graphicx}
\usepackage{caption}
\usepackage{subcaption}
\usepackage{diagbox}
\usepackage[linesnumbered,ruled,vlined]{algorithm2e}
\usepackage{flushend}
\usepackage{mathtools}
\usepackage{amsthm}
\usepackage{amsmath}
\usepackage{verbatim}
\usepackage{pbox}
\usepackage{nth}
\usepackage[linesnumbered,ruled,vlined]{algorithm2e}

\SetCommentSty{mycommfont}
\usepackage{algorithmic}
\usepackage{multirow}
\begin{document}
\date{}
\title{\Large \bf Towards Stochastically Optimizing Data Computing Flows}
\author{
{\rm Farshid Farhat, Diman Zad Tootaghaj, Mohammad Arjomand}\\
The Pennsylvania State University\\
email:{\{fuf111,dxz149,mxa51\} @cse.psu.edu}
}
\maketitle
\thispagestyle{plain}
\pagestyle{plain}
\subsection*{Abstract}
With rapid growth in the amount of unstructured data produced by memory-intensive applications, large scale data analytics has recently attracted increasing interest. Processing, managing and analyzing this huge amount of data poses several challenges in cloud and data center computing domain. Especially, conventional frameworks for distributed data analytics are based on the assumption of homogeneity and non-stochastic distribution of different data-processing nodes. The paper argues the fundamental limiting factors for scaling big data computation. It is shown that as the number of series and parallel computing servers increase, the tail (mean and variance) of the job execution time increase. We will first propose a model to predict the response time of highly distributed processing tasks and then propose a new practical computational algorithm to optimize the response time.
\vspace{-1\baselineskip} 
\section{Introduction}\vspace{-1\baselineskip}
Huge and growing dataset sizes has lead to the emerging field of \textit{Big Data analytics}. As we enter the "petabyte Age",  and since data sizes are growing much faster than the processing speeds of computing devices, managing and processing large dataset sizes, scattered over multiple nodes, becomes more challenging \cite{anderson2008petabyte, mayer2013big}. In order to process such huge datasets, data computation is usually done in parallel and over multiple nodes. Mapreduce, \cite{dean2008mapreduce} and its implementations like Hadoop \cite{hadoop} and Dryad \cite{isard2007dryad} are examples of such distributed programming frameworks for distributed processing of very large dataset sizes.  
These programming frameworks have been optimized for homogeneous hardware architecture, however there exists heterogeneity in the system. Resource sharing which may lead to resource contention, network and memory bandwidth queuing delays, power constraints, inherent heterogeneous workload and computing/data nodes are all different factors which may result in different service rates for different servers in large scale distributed computing. Therefore, we are not able to get the desired performance from the current system.\\
\begin{figure}[!tb]
\centering
\includegraphics[height=3in, width=3.3in]{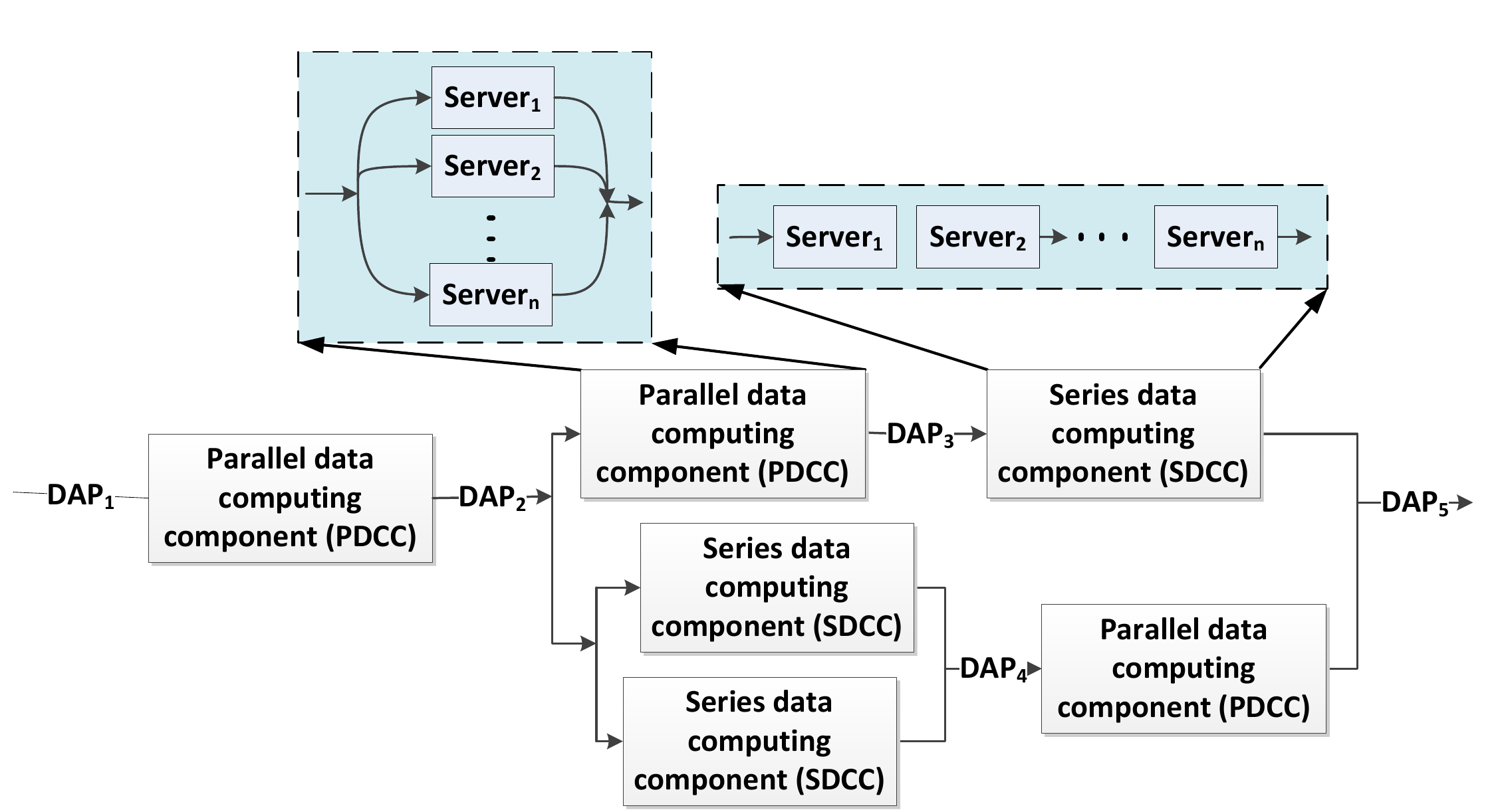}
\caption{\label{fig:DataFlow} An example of dataflow with both series and parallel components.}
\vspace{-1\baselineskip} 
\end{figure}
\indent Distributed jobs have to confront with performance variation which is caused by resource contention, network contention or heterogeneity in workload or computing nodes. The stochastic performance variation (also known as stragglers) can cause as high as 100x performance degradation on overall job computation time \cite{dean2013tail,ananthanarayanan2010reining}. In general, as the number of parallel/series processing components increases, the mean and variance of total job completion time also increases \cite{dean2013tail}.\\
\begin{figure*}[tb]
        \centering
        \begin{subfigure}[b]{0.5\textwidth} 
                \includegraphics[height=1.5in, width=3in]{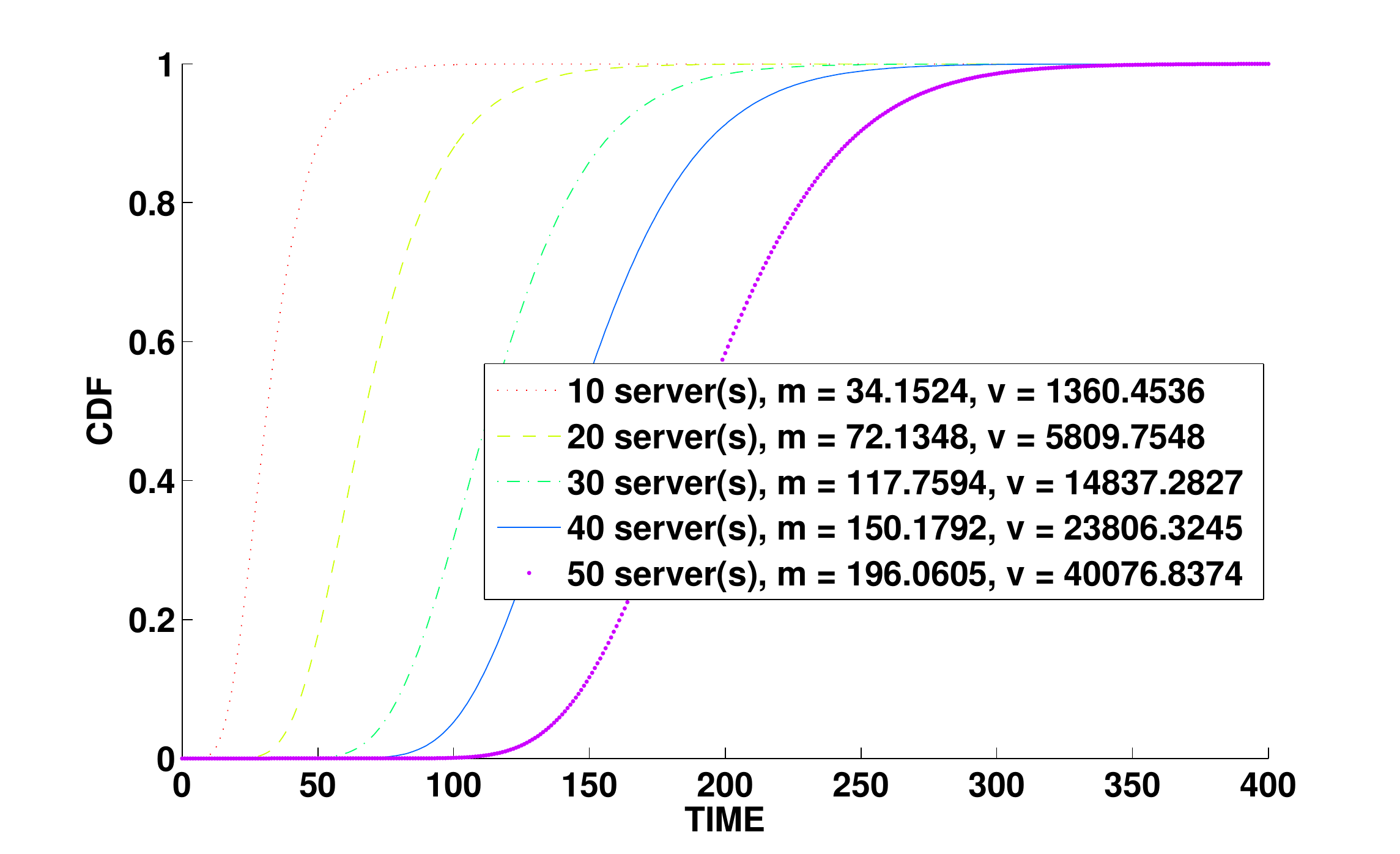}
                \caption{\label{fig:distribution} CDF.} 
        \end{subfigure}%
        ~ 
        \begin{subfigure}[b]{0.5\textwidth}
                \includegraphics[height=1.5in, width=3in]{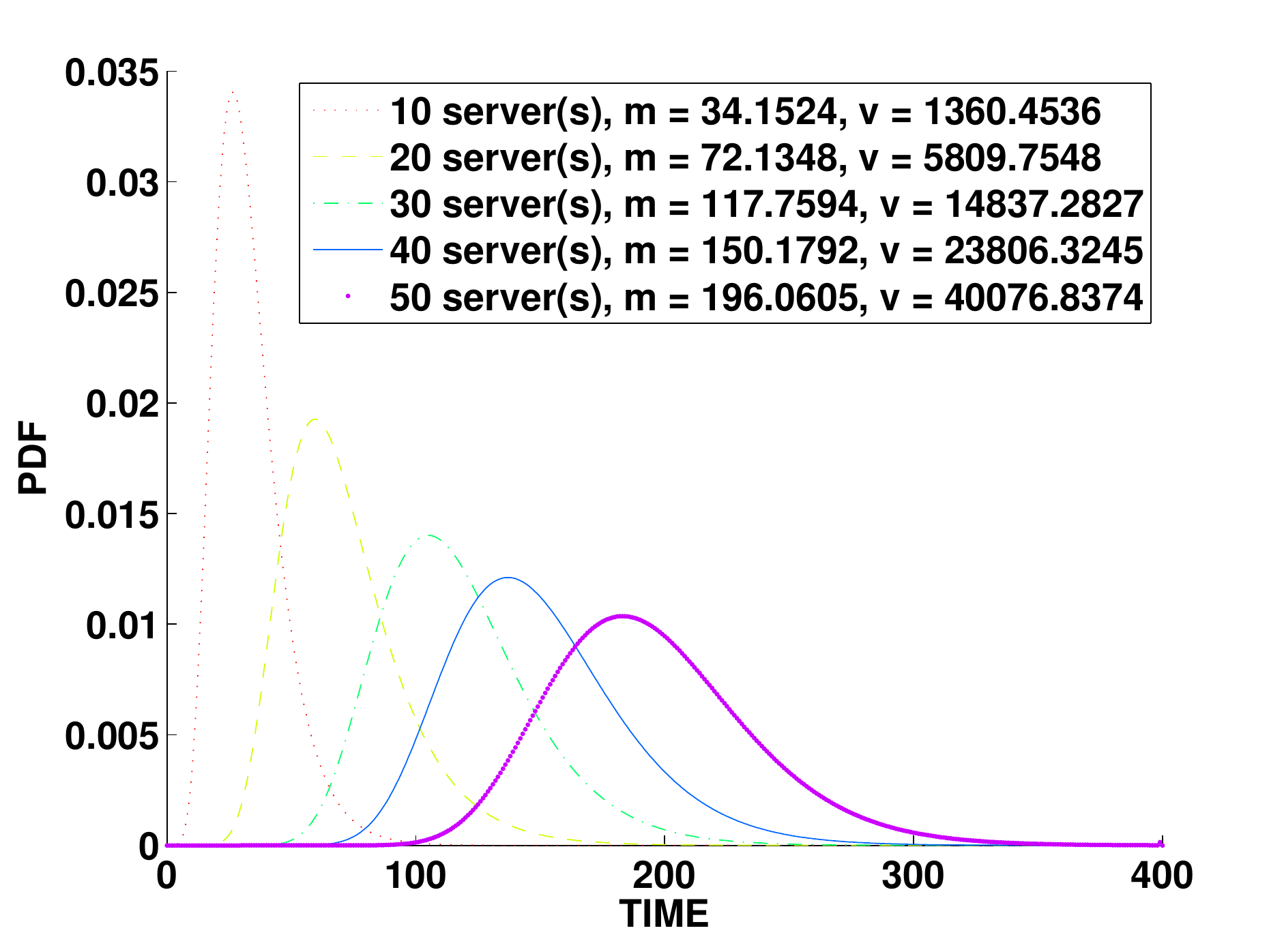}
                \caption{\label{fig:distribution_N} PDF.}
        \end{subfigure}   
 \caption{Job execution time distribution function of 10-50 serial servers.}\label{fig:Distribution}
\vspace{-1\baselineskip} 
\end{figure*}
\begin{figure*}[tb]
        \centering
        \begin{subfigure}[b]{0.5\textwidth} 
                \includegraphics[height=1.5in, width=3in]{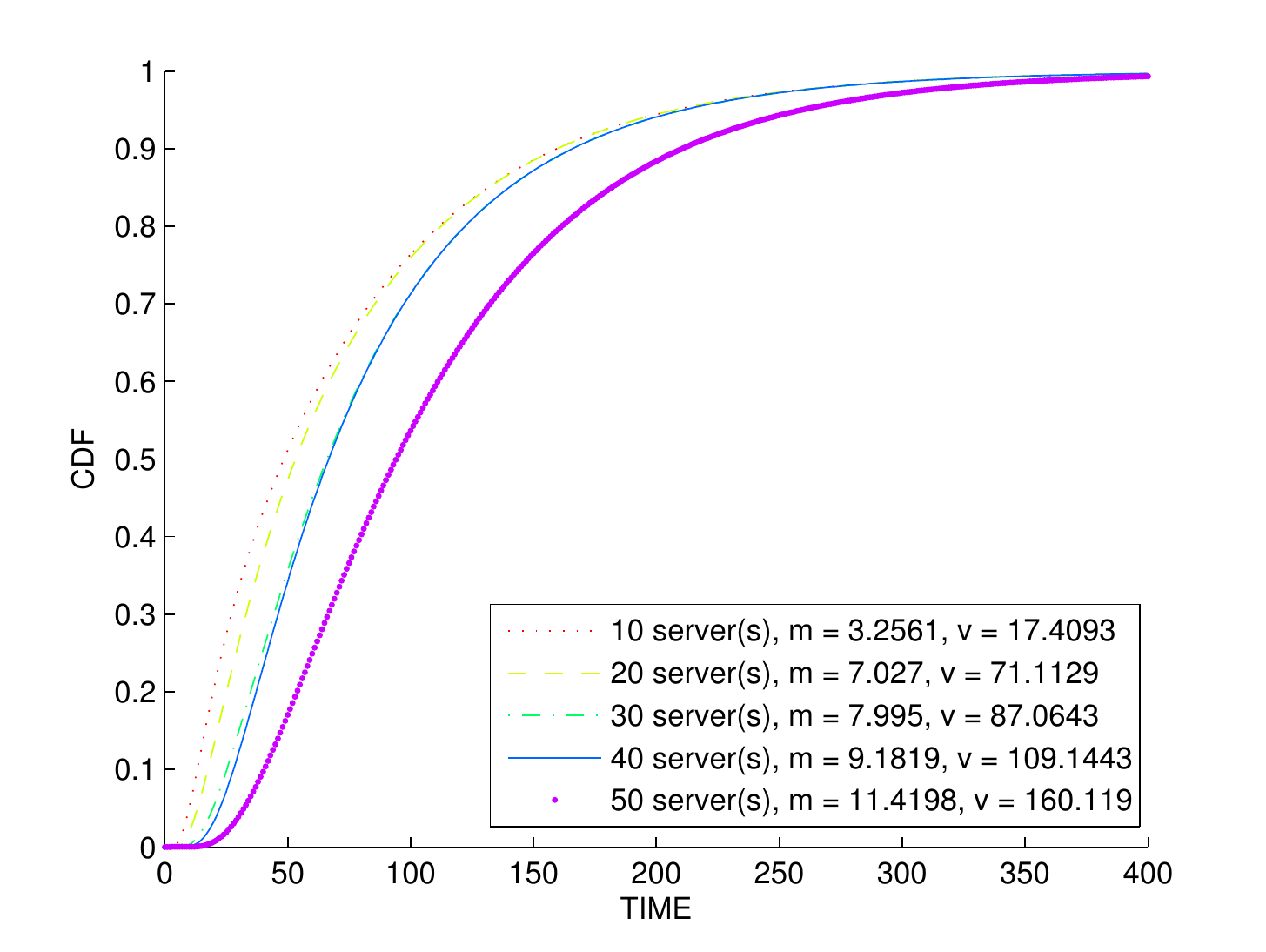}
                \caption{\label{fig:distribution_parallel} CDF.}
        \end{subfigure}%
        ~ 
        \begin{subfigure}[b]{0.5\textwidth}
                \includegraphics[height=1.5in, width=3in]{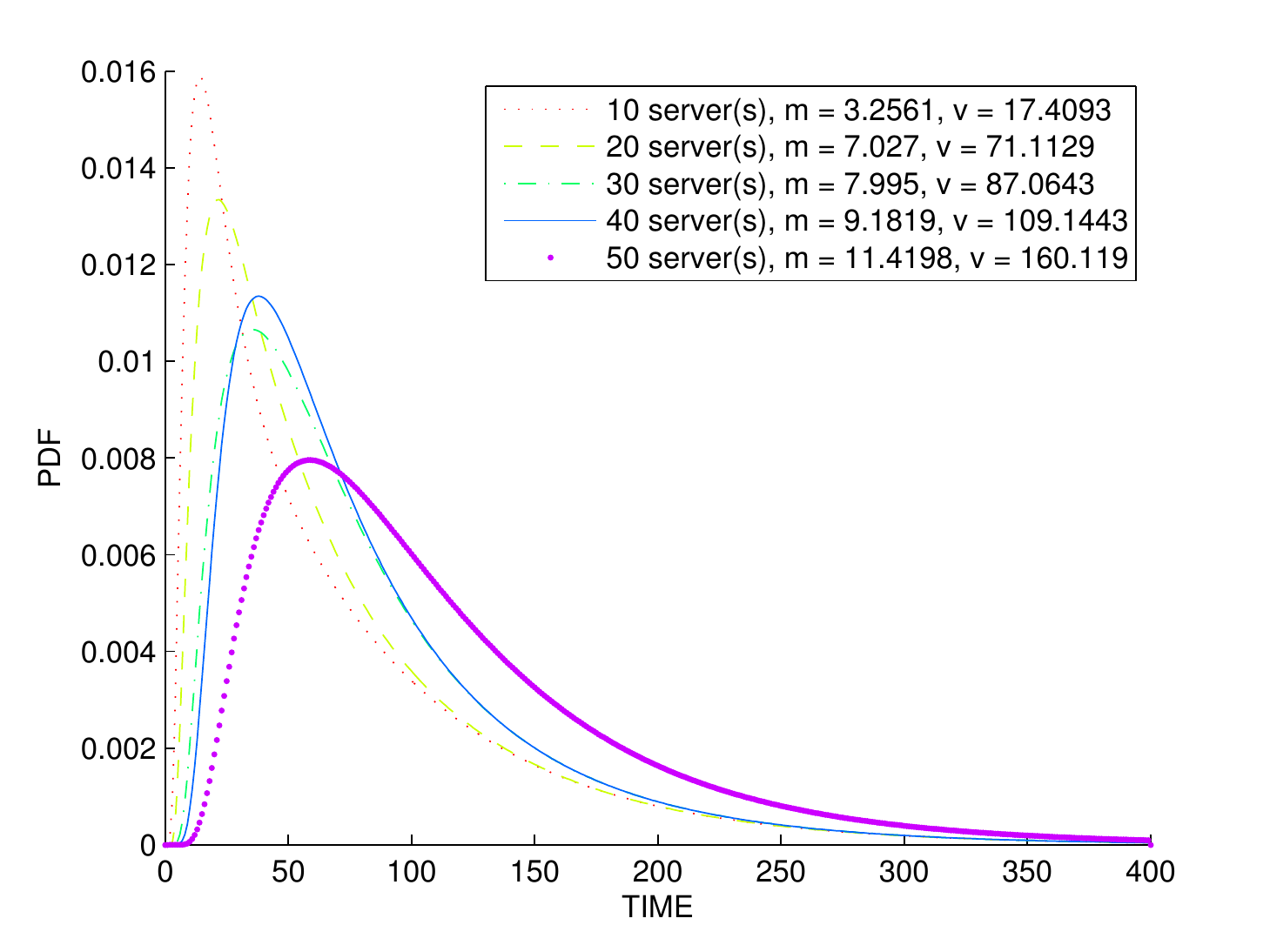}
                \caption{\label{fig:distribution_N_parallel} PDF.}
        \end{subfigure}   
 \caption{Job execution time distribution function of 10-50 parallel servers.}\label{fig:Distribution_parallel}
\vspace{-1\baselineskip} 
\end{figure*}   
\indent While prior works \cite{ahmad2012tarazu, delimitrou2013paragon, yeo2011using, zaharia2008improving, boutaba2012cloud, reiss2012heterogeneity, kurve2013agent, kotobi2015data} have addressed such heterogeneity and stochastic response time of servers in data centers, non-stochastic (deterministic) assumption is still widely used in commercial data centers due to 1) limited mathematical modeling available for such stochastic behavior and 2) lack of a practical method to perform stochastic analysis to optimize job scheduling based on the stochastic system assumption.\\
\indent Furthermore, each Mapreduce job is a smaller part of the chain for big data analysis. For example, Google uses a chain of Mapreduce operations for indexing. Figure~\ref{fig:DataFlow} shows an example dataflow chain which consist of series and parallel components. In this paper's figures, "$m$" stands for mean and "$v$" stands for variance. Each fork/join point in Figure~\ref{fig:DataFlow} is equivalent to a data node that we call it \textit{Data Access Point (DAP)}, which can have a different stochastic distribution for inter-arrival times. \\
\indent As shown in Figure~\ref{fig:DataFlow}, the incoming data is first accessed by the \textit{Data Computing Component (DCC)} at the data access point,  called DAP. Then, there may be a \textit{Serial/Parallel Data Computing Component}, called \textit{SDCC} or \textit{PDCC}, to do the necessary process. Then, the processed data may flow in other branches of the job workflow.\\
\indent Figures~\ref{fig:distribution} and ~\ref{fig:distribution_N} plot the total end-to-end service time distributions of 10 to 50 server nodes that each have exponential distribution when they run in series. Similarly, figures~\ref{fig:distribution_parallel} and ~\ref{fig:distribution_N_parallel} show the total delay distributions of 10 to 50 parallel servers with exponential service time. We will explain the figure and our serial computing model in more details in Section~\ref{subsection:Serial_Data_Computing_Model}. The first figure shows the cumulative distribution function and the second one shows the probability distribution function. These figures illustrate how the mean and variance of job service time of a series of exponential distribution increase as the number of series components increase. Therefore, the tail of the graph increases as the number of serial or parallel components increase, even if each job has an exponential distribution, the whole system has a long tail distribution which is a limiting factor for scaling big data computation. Mapreduce, is used for batch processing of large-scale datasets. It has been shown that query-based jobs also face the same problem \cite{dean2013tail, ananthanarayanan2012let}.\\
\indent \textbf{Problem definition:} Consider a system with $M$ heterogeneous servers with different service rates that (collectively) need to process a data workflow (e.g., Figure~\ref{fig:DataFlow}). The objective of resource allocation and task scheduling is to place the servers inside DCCs and distribute the tasks by adjusting the rates of DAPs for better performance. For example, let us assume that we have a 4 servers available at time $t$ with service rates of 2, 4, 6, and 8. Then, the questions we want to answer are which servers should compute which part of the workflow and how much work we can schedule for each of them?\\
\indent \textbf{A brief description of the proposed model:} We propose to monitor each data access point (DAP), and based on the monitored distribution of each DAP, distribute the jobs such that the total execution time would be minimized. It is shown that any distributed job can be modeled as series and parallel servers \cite{GoogleCloud, zhang2012optimizing}. We find an analytical model for service time distribution of the parallel and series servers and frequently monitor the distribution over time. Based on the monitored distribution we find optimum scheduling algorithm to distribute the jobs among servers. To best of our knowledge, this is the first paper that proposes to perform Job scheduling using the stochastic variation of service time of different servers and the dataflow graph.
\vspace{-1\baselineskip} 
\section{Model Description} \label{section:Model_description} \vspace{-1\baselineskip} 
\indent A job is a process that needs to be executed in a specified time. Executing a large dataset in the cloud environment or in a data center may require a sequence of series or parallel data processing phases, where each phase can be run sequentially or in parallel. In fact, in most cases. the workflow of the job is based on its algorithm. Note however that, the amount of resources provided to the job should be allocated and scheduled by the system administrator dynamically.\\
\indent Previous works have used deterministic (not stochastic) performance metrics \cite{ahmad2012tarazu, delimitrou2013paragon, yeo2011using} for each phase and optimized the end-to-end performance of the workflow using computing/processing power of different components in the path.
There exists no stochastic modeling of job workflow in the cloud or data center environment. Our proposed model considers, an execution time delay for each server node as well as a waiting time distribution to perform the task. It is assumed that a server is a \textit{queue}, where tasks come for service with a specific service rate. \\
\indent Most of the servers' waiting time distribution can be modeled as a short/long delayed tail distribution. So the underlying distribution is exponential/pareto or a combination of them (known as multi-modal distribution). The main distributions which have been modeled in this paper are listed in Table~\ref{table:PDF}. Using real data center service time distributions it is shown in \cite{ananthanarayanan2010reining, ren2013hadoop, farhatstochastic, farhat2015stochastic, farhat2014modeling, tootaghaj2015evaluating, diman2015evaluating} that the cumulative distribution function of service time of servers can be modeled as six different distributions listed in Table ~\ref{table:PDF}. The delayed step function $U(t-T_i)$ specifies the minimum amount of time to complete a task.
\begin{table}[!htb]\scriptsize
\centering
\caption{\label{table:PDF} Cumulative distribution functions of service time distributions used in the paper.}
\begin{tabular}{|p{1.3in}||p{1.5in}|} 
\hline Distribution name & Cumulative distribution function \\
\hline Delayed exponential & $F_{{de}_i}(t)=(1- \alpha e^{-{\lambda}_i (t-T_i)}) U(t-T_i)$\\
\hline Delayed pareto & $F_{{dp}_i}(t)=(1- \alpha e^{-{\lambda}_i (ln(t+1)-T_i)}) U(t-T_i)$ \\
\hline Multi-modal delayed exponential (DE) & \pbox{60cm}{$F_{mmde}(t)= \sum\limits_{i=1}^n p_i F_{de}(t);$\\ Where $\; \; \; \; \; \;  \sum\limits_{i=1}^n p_i = 1 \; \; \; \; \; \; \forall i \; \; \; p_i \geq 0$}  \\
\hline Multi-modal delayed pareto (DP) & \pbox{60cm}{$F_{mmdp}(t)= \sum\limits_{i=1}^n p_i F_{dp}(t);$\\ Where$ \; \; \; \; \; \;  \sum\limits_{i=1}^n p_i = 1 \; \; \; \; \; \; \forall i \; \; \; p_i \geq 0$} \\
\hline Delayed tail & \pbox{60cm}{$F_{{d}_i}(t)=(1- \alpha e^{-{\lambda}_i (m(t)-T_i)}) U(t-T_i)$ \\  Where $m(t)$ is monotonically increasing} \\
\hline Multi-modal delayed tail & \pbox{60cm}{$F_{mmd}(t)= \sum\limits_{i=1}^n p_i F_{d_i}(t);$\\Where  $\; \; \; \; \; \; \sum\limits_{i=1}^n p_i = 1 \; \; \; \; \; \; \forall i \; \; \; p_i \geq 0$} \\
\hline
\end{tabular}
\end{table}

\vspace{-1\baselineskip} 
\subsection{Serial Data Computing Model}\label{subsection:Serial_Data_Computing_Model} 
\begin{figure}
\centering
\includegraphics[height=0.5in, width=3in]{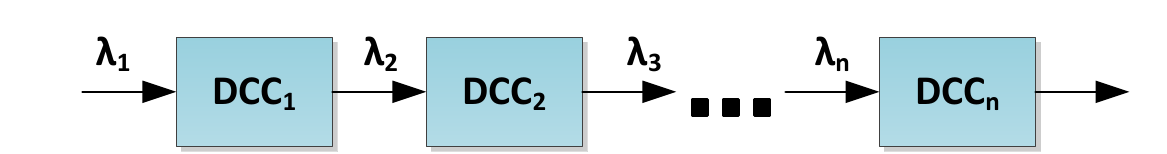}
\caption{\label{fig:serial} An example of serial DCCs.}
\vspace{-1\baselineskip} 
\end{figure}

\vspace{-1\baselineskip} 
In a serial data computing component, the data passes through a set of DCCs sequentially till it gets out from the last one. As shown in Figure~\ref{fig:serial}, the sequential DCCs form a tandem queue with response times of $X_1$, $X_2$, ..., $X_n$.
 The end-to-end response time of tandem queue of DCCs is $\sum_i X_i$, and consequently the distribution of end-to-end response time would be the convolution of the residual distribution.
We have: 
\newcommand{\Conv}{\mathop{\scalebox{2.5}{\raisebox{-0.2ex}{$\ast$}}}}%
\begin{equation}
f_{X_1 + X_2 + ... + X_n}(t) = f_{X_1}(t)* f_{X_2} * ...* f_{X_n}(t)= \Conv^{i=1}_{n} (f_{X_i}(t)).
\end{equation}
For example, the sum of two exponential distributions $F_1(t) = 1- e^{-\lambda_1 t}$ and $F_2(t)= 1- e^{-\lambda_2 t}$ is 
\begin{equation}
F(t)=F_1(t) * F_2(t)= 1- \frac{\lambda_2}{\lambda_2 - \lambda_1} e^{- \lambda_1 t} +\frac{\lambda_1}{\lambda_2 - \lambda_1} e^{- \lambda_2 t} 
\end{equation}
The effect of job serialization was shown in Figure~\ref{fig:Distribution}. As the number of serial servers increases, the probability distribution function of end-to-end response time shifts to right, i.e., the mean and variance of the job completion increases.  
\vspace{-1\baselineskip} 
\subsection{Parallel Data Computing Model} 
\begin{figure}[!t]
\centering
\includegraphics[height=1in, width=2.5in]{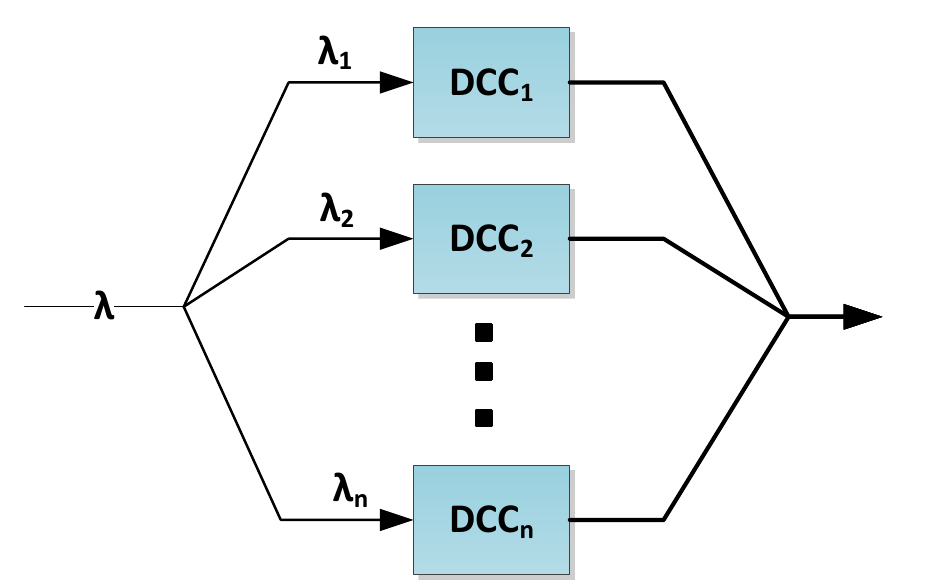}
\caption{\label{fig:parallel} An example of parallel DCCs.}
\vspace{-1\baselineskip} 
\end{figure}

\vspace{-1\baselineskip} 
In a parallel data computing component, the data is partitioned and sent through a set of DCCs in parallel; and the task is completed when the last DCC sends its result to the next joint data access point, as shown in Figure~\ref{fig:parallel}. The end-to-end response time (between $\mathrm{DAP_{1}}$ and $\mathrm{DAP_{2}}$ in the Figure) of fork-join servers is the $max (X_1, X_2, . . ., X_n)$, where $X_i$ is the response time of i$^{\textnormal{th}}$ DCC. Therefore, we have:
\begin{align}
F_{max(x_1, ..., x_n)} (t) = P(max(X_1, ..., X_n) \leq t)= \nonumber\\ 
P(X_1 \leq t) P(X_2 \leq t) ... P(X_n \leq t)= \nonumber\\ F_{X_1} (t) .... F_{X_n} (t) = \prod\limits_{i=1}^n F_{X_i} (t)
\end{align}
For example, the maximum of two exponential distributions $F_1(t)= 1 - e^{-\lambda_1 t} $ and $F_2(t)= 1 - e^{-\lambda_2 t} $ is 
\begin{equation}
F(t)= F_1(t) F_2(t)=(1 - e^{-\lambda_1 t})(1 - e^{-\lambda_2 t}) 
\end{equation}
As shown in Figures~\ref{fig:distribution_parallel} and~\ref{fig:distribution_N_parallel}, when the number of parallel servers increases, the mean and variance of job completion time elevates and the PDF moves to the right. However, the magnitude of parallel effect is lower than the serial effect. This result also verifies the effect mentioned in \cite{dean2013tail}. 
\vspace{-1\baselineskip} 
\section{Optimization of Job Workflow} 
\begin{figure}[!tb]
\centering
\includegraphics[height=1in, width=3.2in]{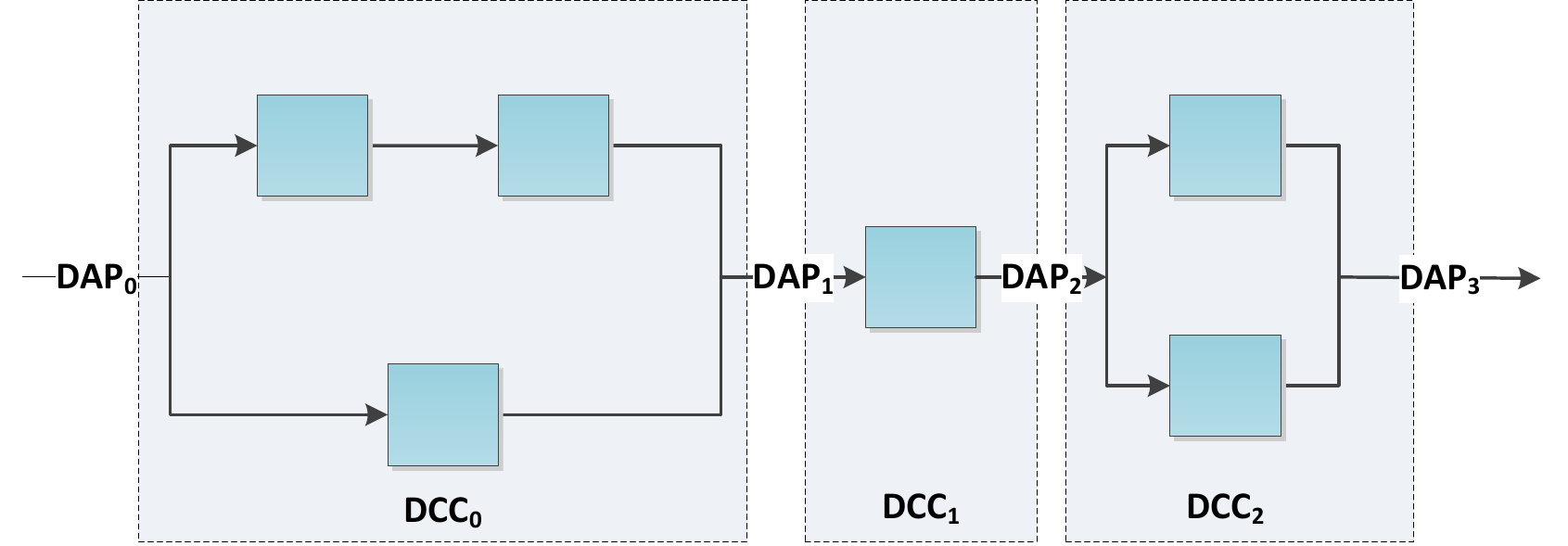}
\caption{\label{fig:Building_Blocks} Logical view of a job workflow including serial and parallel DCCs.}
\vspace{-1\baselineskip} 
\end{figure}

\vspace{-1\baselineskip} 
\indent From the administrative point of view, we want to know the end-to-end distribution of the system and we want to optimize the end-to-end response time or maximize throughput.\\
\indent The goal is to find a resource allocation (placing servers inside DCCs) and task scheduling (adjusting rates of DAPs) for better performance. Here, we aim for throughput or response time; however, our optimization strategy can also be used for other objective functions as well. The main lemma (not proven here because of space limit) is:
\newtheorem{lemma}{Lemma}
\begin{lemma}
\label{Optimization}
The global optimization scheme can be divided and conquered; i.e., it can be decomposed into smaller optimization problems on serial and parallel data computation components.
\end{lemma}
\indent To have a better throughput, the waiting time of all serial component must be minimum and the same. In other words, we desire to minimize the delay of the SDCC which has the highest delay.\\
\indent To have a better response time, the parallel DCCs must have the same lowest statistical moment, such as first moment (mean) or second degree moment (variance) or equivalently, we desire to minimize the longest delay of parallel components. The necessary information to manage job workflow is the performance distribution of each server which is gradually updated over the time. \\
\indent In order to implement data computing flow management, we focus on minimizing response time which is the dual optimization of maximizing the throughput. 
Finding the best response time of the workflow leads to the problem of finding the best response time of its sequential building blocks. For example, Figure~\ref{fig:Building_Blocks} shows a logical view of a job workflow captured by an iterating algorithm in big data analysis. Therefore, the minimization of the response time from $\mathrm{DAP_{0}}$ to $\mathrm{DAP_{3}}$ is equivalent to the minimization of the response times of $\mathrm{DCC_{0}}$, $\mathrm{DCC_{1}}$ and $\mathrm{DCC_{2}}$. Now, if $\mathrm{DCC_{i}}$ corresponds to a SDCC (as in $\mathrm{DCC_{1}}$), we use SDCC optimization algorithm; and if $\mathrm{DCC_{i}}$ corresponds to a PDCC, as in $\mathrm{DCC_{0}}$ and $\mathrm{DCC_{2}}$, we employ PDCC optimization algorithm recursively. \footnote{Note that there may be other $DCC$s inside a $DCC$ as well.}\\
\begin{algorithm}[!htb]\scriptsize
\DontPrintSemicolon 
\KwIn{Available server's response time distribution (RES\textunderscore Array), and data arrival rates (amount of task) in each DAP (DCCRES\textunderscore Array).}
\KwOut{Server (resource) allocation and data rate (task) scheduling.}

Sort available servers based on expected values of their response time distribution in descending order in an array (RES\textunderscore Array)\\

Sort DCCs based on their data arrival rates in ascending order in an array (DCC\textunderscore Array).

\tcc{Allocate the sorted RES\textunderscore Array to the sorted DCC\textunderscore Array respectively.}
\begin{algorithmic}
\FOR {$i = 1 \to size(DCC\textunderscore Array)$}
\IF{ DCC\textunderscore Array[i] is a single queue)}
\STATE Place RES\textunderscore Array[1] in DCC\textunderscore Array[i]
\STATE Remove the head of RES\textunderscore Array 
\ELSIF {DCC\textunderscore Array[i] == SDCC}
\STATE $SDCC\textunderscore Allocate(RES\textunderscore Array, DCC\textunderscore Array[i])$
\ELSIF {DCC\textunderscore Array[i] == PDCC}
\STATE $PDCC\textunderscore Allocate(RES\textunderscore Array, DCC\textunderscore Array[i])$
\ENDIF
\ENDFOR
\end{algorithmic}
\caption{SDCC\textunderscore allocate Algorithm.}
\label{algo:SDCC_allocate}
\end{algorithm}
\indent As a result of SDCC\textunderscore allocate algorithm, the faster servers are placed into the DCC with higher data arrival rates. For parallel building blocks we run Algorithm~\ref{algo:PDCC_allocate} recursively.\\
\begin{algorithm}[!htb]\scriptsize
\DontPrintSemicolon 
\KwIn{Sorted available servers (RES\textunderscore Array) based on response time distribution in descending order and data arrival rates (amount of task) in each DAP.}
\KwOut{Server (resource) allocation and data rate (task) scheduling.}

\begin{algorithmic}
\IF {$\lambda_i$s are known}
\STATE{Sort DCCs list based on their $\lambda_i$s as in DCC\textunderscore Array}
\STATE Allocate DCC\textunderscore Array with RES\textunderscore Array respectively (Allocate(RES\textunderscore Array, DCC\textunderscore Array).
\ENDIF
\IF {$\lambda_i$s are unknown and only their sum ($\lambda$) is known}
\STATE {Sort DCCs based on the number of internal DAPs in descending order as an array (DCC\textunderscore Array)} 
\FOR {$i = 1 \to size(DCC\textunderscore Array)$}
\IF{ (DCC\textunderscore Array[i] is a single queue)}
\STATE Place RES\textunderscore Array[1] in DCC\textunderscore Array[i]
\STATE Remove the head of RES\textunderscore Array 
\ELSIF {DCC\textunderscore Array[i] == SDCC}
\STATE $SDCC\textunderscore Allocate(RES\textunderscore Array, DCC\textunderscore Array[i])$
\ELSIF {DCC\textunderscore Array[i] == PDCC}
\STATE $PDCC\textunderscore Allocate(RES\textunderscore Array, DCC\textunderscore Array[i])$
\ENDIF
\ENDFOR
\ENDIF
\end{algorithmic}

\textbf{Rate scheduling:} By solving the following equilibrium, we can find the data arrival rates (amount of tasks) sent for each DCC:\\
\begin{center}
$\lambda =\sum_{i=1}^{n} \lambda_i,\nonumber$ \\
$\lambda_1 RT_{DCC_1}= \lambda_2 RT_{DCC_2}= ...= \lambda_n RT_{DCC_n}$\\
Where $RT$ is the response time of $DCC_i$ and $\lambda_i$ is the data arrival rate to $DCC_i$ as shown in Figure~\ref{fig:parallel}
\end{center}
\caption{PDCC\textunderscore allocate Algorithm.}
\label{algo:PDCC_allocate}
\vspace{0\baselineskip}
\end{algorithm}
\indent To perform data computing flow management, we propose Algorithm~\ref{algo:b}.
\begin{algorithm}[!htb]\scriptsize
\DontPrintSemicolon 
\KwIn{Performance distribution of each server and logical graph of the job.}
\KwOut{Distribution of workflows.}
Get the logical workflow of a job based on an iterating algorithm (DCC\textunderscore Array).

Amount of work to be done by a server, i.e. arrival rate of task to each server (RES\textunderscore Array).

Compute power of a server, i.e. recent waiting time distribution of each server.

SDDC\textunderscore allocate(RES\textunderscore Array, DCC\textunderscore Array)
\caption{Management of data computing flows.}
\label{algo:b}
\vspace{0\baselineskip}
\end{algorithm}
It is assumed that the logical graph of the job workflow is known using a computational algorithm (out of the scope of this paper). Using Algorithm~\ref{algo:b}, the DCCs flow is extracted. If there exists a SDCC or PDCC, the algorithm calls Algorithm~\ref{algo:SDCC_allocate} or~\ref{algo:PDCC_allocate}, respectively, to perform resource allocation/task scheduling.\\

\begin{figure*}[tb]
        \centering
        \begin{subfigure}[b]{0.5\textwidth} 
                \includegraphics[height=1.5in, width=3in]{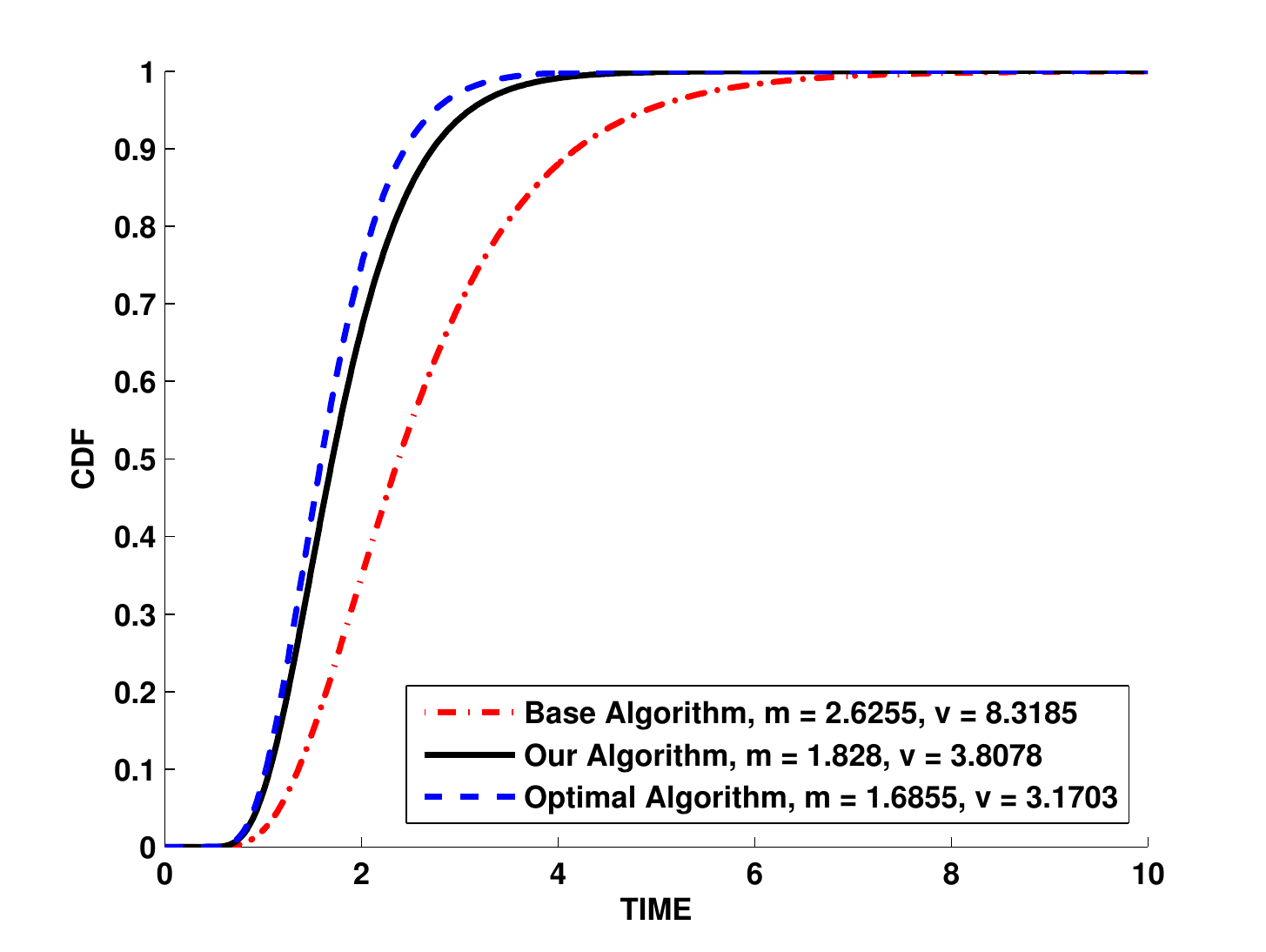}
                \caption{\label{fig:optimized_parallel} CDF.}
        \end{subfigure}%
        ~ 
        \begin{subfigure}[b]{0.5\textwidth}
                \includegraphics[height=1.5in, width=3in]{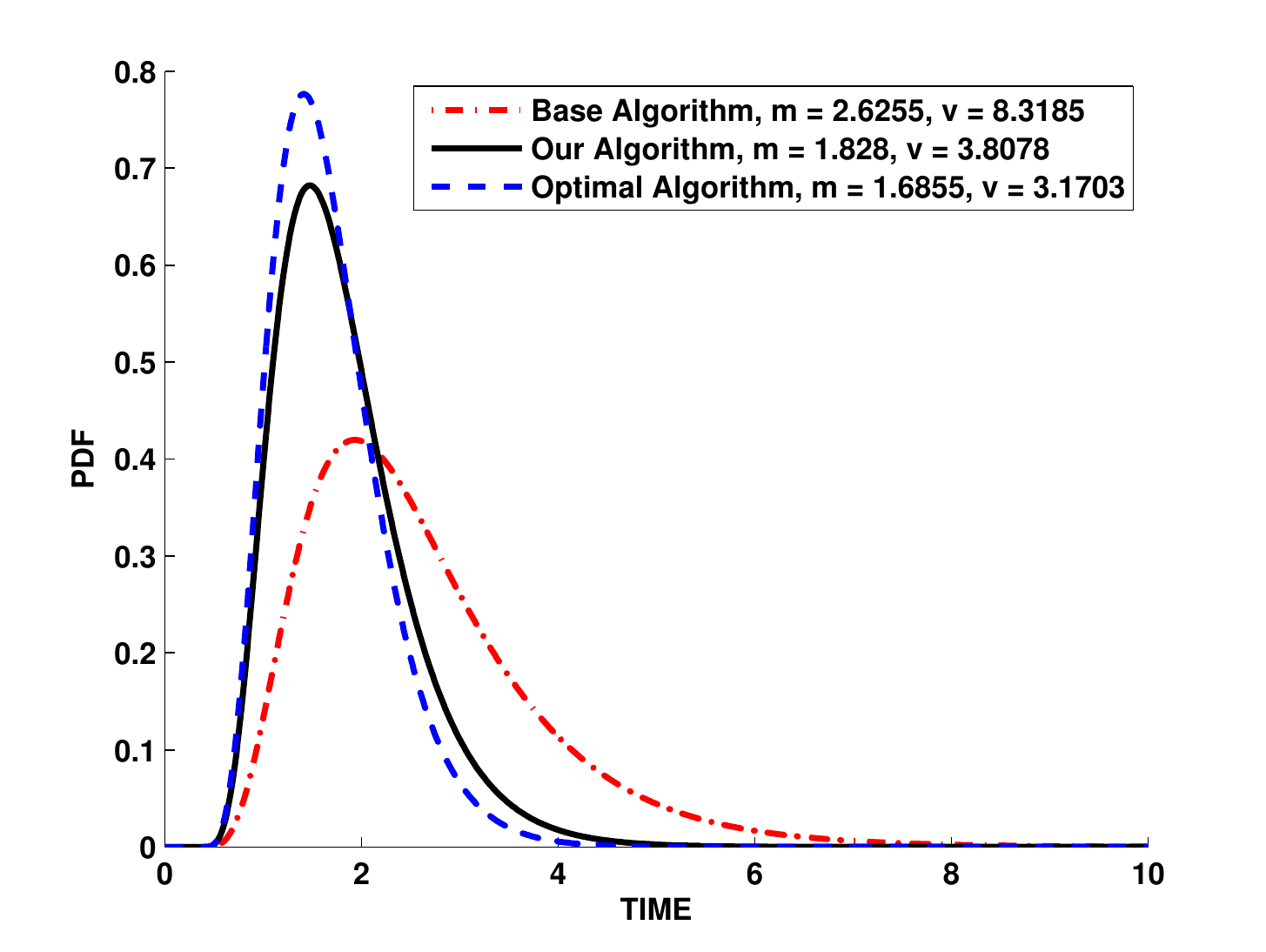}
                \caption{\label{fig:optimized_N_parallel} PDF.}
        \end{subfigure}   
 \caption{Comparison of response time distribution between the baseline, optimal and our scheme.}\label{fig:Optimized}
\vspace{-0.5\baselineskip} 
\end{figure*}

\indent As shown in Figure~\ref{fig:Optimized}, we tested our proposed algorithms, the optimal method, and the heuristic baseline algorithm, for the workflow depicted in Figure~\ref{fig:Building_Blocks}, where DAPs rates were $\lambda_{DAP_0}=8$, $\lambda_{DAP_1}=4$, and $\lambda_{DAP_2}=2$, and available servers have the service rates of 9, 8, 7, 6, 5 and 4. We used different CDFs from Table~\ref{table:PDF} and the results is valid for all 6 distribution functions. In the heuristic baseline algorithm we use the algorithm that allocates SDCCs sooner than PDCCs and for parallel components, jobs are distributed using the equilibrium equation in Algorithm~\ref{algo:PDCC_allocate}. Heuristic baseline algorithm first allocates better servers to SDCCs (as they become intuitively bottleneck servers), and then allocates PDDCs. Optimal method chooses the best allocation of servers (using exhaustive search over all possible cases) to DCCs and uses optimal task scheduling for PDCCs. To be fair, we used the optimal task scheduling for the heuristic baseline algorithm (however in real systems it may not be true if the scheduler assigns same amount of work to all servers with homogeneous assumption), but our proposed algorithm is as mentioned before.\\

\begin{table*}[tb]
\centering
\caption{\label{table:results} Simulation results for different distribution functions listed in Table~\ref{table:PDF}.}
\begin{tabular}{|p{2cm}|l|l|l|l|l|l|l|l|}
    \hline
   \multirow{1}{*}{\parbox{1.2cm}{\textbf{Performance metric}}}
  & \multicolumn{4}{|c|}{\textbf{mean}}  & \multicolumn{4}{|c|}{\textbf{Variance}}  \\
    \cline{2-9}
     & our approcah & optimal & baseline & improvement & our approcah & optimal & baseline & improvement  \\ 
    \hline
   \textbf{Scenario 1} & 1.828 & 1.6855 & 2.6255 & 30.38$\%$ & 3.8078 & 3.1703 & 8.3185 & 54$\%$  \\      
    \hline
    \textbf{Scenario 2} & 2.8637 & 2.5251 & 5.4125 & 47.1$\%$ & 11.4057 & 8.3924 & 39.3003 & 71$\%$  \\ 
    \hline
    \textbf{Scenario 3} & 2.5038 & 2.09 & 4.4089 & 43.2$\%$ & 8.8881 & 5.4871 & 27.741 & 68$\%$  \\ 
    \hline
\end{tabular}
\end{table*}

\indent Table~\ref{table:results} shows the results (mean and variance) of multiple possible scenarios, when the available servers' service time distribution are delayed exponential, delayed pareto, or mix of them. Based on these promising results, we can always get improvement for mean and variance of the system response time over the baseline allocation/scheduling algorithm with a little gap from the optimal choice.\\

\section{Concluding Remarks}\vspace{-1\baselineskip} 
In this study, we conducted an analytic study of different dataflows in distributed analysis to monitor data access points and based on service time distribution of servers at each data access point we propose a scheduling algorithm to minimize the total execution time.\\
\indent While we made our proposed approach simple, more complicated dynamic task scheduling could be used with higher complexity. In fact, the software and hardware concepts in the cloud were modeled or mathematically mapped as queuing theoretic building blocks, while most of such mapping description was skipped as well as some fundamental lemmas.\\
\indent Knowing the data arrival rates and service rates before execution may have the overhead to run the framework dynamically. This approach forms a bridge between machine learning schemes and analytical schemes. We believe that our proposed approach has lower overhead than the machine learning schemes for managing large number of servers. However, practical issues of the approach have yet to be well investigated.
\vspace{-1\baselineskip} 
\bibliographystyle{unsrt} 
\bibliography{ieeeconf} 
\end{document}